\documentclass{article}
\usepackage{spconf,amsmath,graphicx}

\usepackage{import, xcolor} %
\usepackage{siunitx}
\usepackage{multirow}
\usepackage{tabularx}
\usepackage{url}

\definecolor{LMSred}{rgb}{0.80,0.20,0.20}

\let\normalsize\small\normalsize
\title{ENHANCED IMAGE RECONSTRUCTION FROM QUARTER SAMPLING MEASUREMENTS USING AN ADAPTED VERY DEEP SUPER RESOLUTION NETWORK}
\name{Simon Grosche, Kristian Fischer, Fabian Brand, J\"urgen Seiler, and  Andr\'e Kaup }
\address{Multimedia Communications and Signal Processing\\ Friedrich-Alexander-Univerist\"at Erlangen-N\"urnberg, Cauerstr. 7, 91058 Erlangen, Germany\\ \textit{\small \{simon.grosche, kristian.fischer, fabian.brand, juergen.seiler, andre.kaup\}@fau.de}}

\usepackage[firstpage=true]{background}
\usepackage{hyperref}

\SetBgContents{\parbox{\textwidth}{\footnotesize © 2020 IEEE. Personal use of this material is permitted. Permission from IEEE must be
		obtained for all other uses, in any current or future media, including
		reprinting/republishing this material for advertising or promotional purposes, creating new
		collective works, for resale or redistribution to servers or lists, or reuse of any copyrighted
		component of this work in other works. DOI: \href{https://doi.org/10.1109/ICIP40778.2020.9191047}{10.1109/ICIP40778.2020.9191047.}}}
\SetBgScale{1}
\SetBgAngle{0}
\SetBgPosition{current page.south}
\SetBgVshift{1cm}
\SetBgColor{black}
\SetBgOpacity{1}

\begin{document}
\maketitle
\begin{abstract}
Quarter sampling is a novel sensor concept that enables the acquisition of higher resolution images without increasing the number of pixels. This is achieved by covering three quarters of each pixel of a low-resolution sensor such that only one quadrant of the sensor area of each pixel is sensitive to light. By randomly masking different parts, effectively a non-regular sampling of a higher resolution image is performed.
Combining a properly designed mask and a high-quality reconstruction algorithm, a higher image quality can be achieved than using a low-resolution sensor and  subsequent upsampling.
For the latter case, the image quality can be enhanced using super resolution algorithms. Recently, algorithms based on machine learning such as the Very Deep Super Resolution network (VDSR) proofed to be successful for this task.
In this work, we transfer the concepts of VDSR to the special case of quarter sampling. Besides adapting the network layout to take advantage of the case of quarter sampling, we introduce a novel data augmentation technique enabled by quarter sampling. Altogether, using the quarter sampling sensor, the image quality in terms of PSNR  can be increased by \SI[retain-explicit-plus]{+.67}{dB} for the Urban 100 dataset compared to using a low-resolution sensor with VDSR. %
\end{abstract}

\section{INTRODUCTION}

\label{sec:intro}
\begin{figure}[t!]
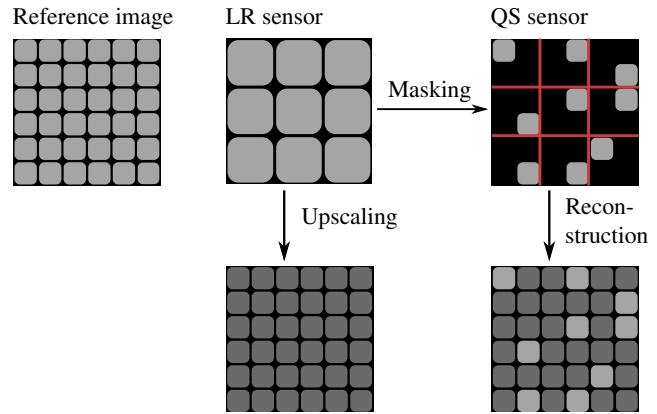
\caption{Illustration of an image acquisition using a low-resolution (LR) sensor and a quarter sampling (QS) sensor. Light gray pixels indicate measured pixels whereas dark gray pixel indicate upscaled/reconstructed pixels.}
\label{fig:LR_vs_quarter}
\end{figure}

Using quarter sampling \cite{Schoberl2011}, the spatial resolution of an imaging sensor can be increased. This is achieved by physically covering three quarters of each pixel of a low-resolution sensor as it is illustrated in Figure\,\ref{fig:LR_vs_quarter}. Effectively, this leads to a non-regular sampling of the image with respect to a higher resolution grid with twice the resolution in both spatial dimensions.
Since the sampling is non-regular, it leads to reduced visible aliasing artifacts conventionally occurring for regular sampling \cite{Dippe1985, Hennenfent2007, Maeda2009}.
The missing pixels need to be reconstructed from the sampled data.
For a proper reconstruction, high-quality reconstruction algorithms such as the frequency selective reconstruction (FSR) \cite{Seiler2015} need to be used in combination with optimized quarter sampling patterns such as those in \cite{Grosche2018}.
FSR has shown to be a successful reconstruction scheme for various inpainting and extrapolation tasks \cite{Herraiz2008, Stehle2006,Seiler2009} and showed best results for non-regular sampling and quarter sampling in~\cite{Schoberl2011,Seiler2015,Grosche2018}.

Another approach to acquire high resolution images without increasing the number of measured pixels is to upscale an image from a low-resolution sensor as it is also shown in Figure\,\ref{fig:LR_vs_quarter}. Unfortunately, using standard interpolation methods leads to a blurred and degraded image. A higher image quality can be achieved using super-resolution algorithms. During the last decade, great progress has been made in this field using sparse representation based approaches \cite{Yang2010} as well as convolutional neural network based approaches \cite{Dong2016}. 
Recently, the Very Deep Super Resolution (VDSR) network \cite{Kim2016} was proposed. Other than previous neural networks, it does not use a low-resolution image as input but enhances an image that was previously upscaled with bicubic interpolation by learning the residuum between the unavailable high resolution image and the upscaled image. Using a rather simple network structure, VDSR achieves a high quality reconstruction especially at sharp edges that otherwise appear blurred by simple interpolation methods such as bicubic upscaling.

Since VDSR is capable of improving the image quality of upscaled images, the question arises whether its concepts can also be used to further enhance images that were reconstructed from a quarter sampling sensor. Though edges are typically reconstructed quite well in that case, other artifacts such as ringing may occur, which could potentially also be removed by learned filters present in VDSR.
In this paper, we transfer the techniques used in VDSR to remove such undesirable artifacts from images reconstructed from quarter sampling sensors while keeping the high-resolution structures made possible by quarter sampling. Besides simply retraining VDSR for the case of quarter sampling in a first step, we further propose an adapted version (\mbox{VDSR-QS}) which incorporates \mbox{knowledge} available for the special case of quarter sampling. Moreover, quarter sampling allows us to exploit a novel degree of freedom for data augmentation which is not accessible for the conventional case of the low-resolution sensor.

This paper is organized as follows: In Section\,\ref{sec:vdsr_and_adapted_vdsr}, we briefly introduce the VDSR network and the corresponding processing chain for the low-resolution sensor. Then, we present the processing chain for a quarter sampling sensor incorporating an adapted VDSR network and introduce a novel technique for data augmentation.
In Section\,\ref{sec:simulation_and_results}, we perform experiments that compare both processing chains and the used adaptations. We furthermore evaluate and discuss the results and provide visual examples.
In Section\,\ref{sec:conclusion}, we summarize the paper and give an outlook to future work.

\section{VDSR IMPLEMENTATIONS FOR LOW-RESOLUTION SENSORS AND QUARTER SAMPLING SENSORS}

\label{sec:vdsr_and_adapted_vdsr}

\subsection{VDSR for a Low-resolution Sensors}

\begin{figure}[t!]
	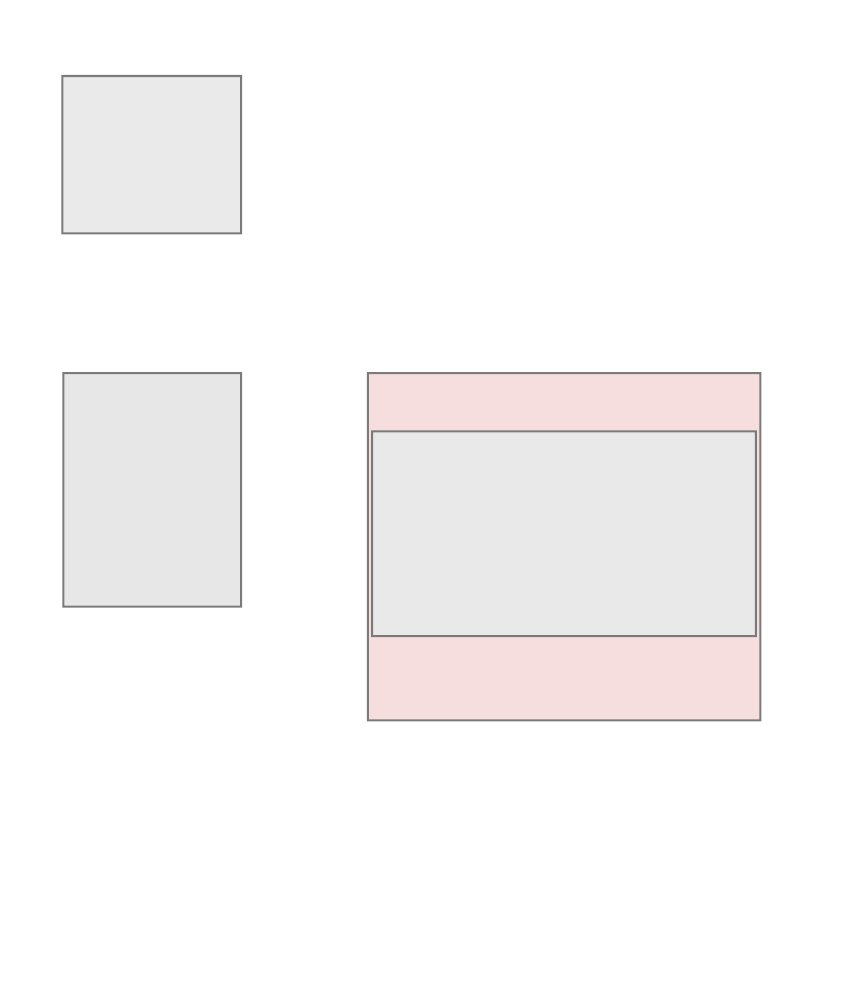
	\caption{Flow graph for sampling and reconstruction for (a) the low-resolution (LR) sensor and (b) the quarter sampling (QS) sensor. For the quarter sampling sensor, a factor $(1-b_{ij})$ is optionally multiplied to the residual output of the neural network as indicated by a red color. }
	\label{fig:fsr_plus_vdsr_network_skizze}
\end{figure}

For a low-resolution sensor as it is shown in Figure\,\ref{fig:LR_vs_quarter}, the acquisition of the image can be described as the  filtering of a high resolution reference image $f_{ij}$ with a $2{\times}2$ filtering kernel of ones followed by a twofold sub-sampling in both spatial dimensions.
Effectively, each pixel measures the mean value of four high resolution pixels from a $2{\times}2$ neighborhood.
Prior to the application of a super-resolution algorithm such as the VDSR \cite{Kim2016}, the image is upscaled using bicubic interpolation leading to an approximate solution $\hat{f}_{ij}$.
In order to enhance the resolution of the upscaled image, we feed the upscaled image $\hat{f}_{ij}$ into VDSR \cite{Kim2016} being a convolution neural network trained to infer the residual $r_{ij} = f_{ij} - \hat{f}_{ij}$. The resulting image is calculated by summing the input and output of VDSR, i.e.,  
\begin{align*}
	\tilde{f}_{ij} = \hat{f}_{ij} + r_{ij}.
\end{align*}
The full processing chain for the case of the low-resolution sensor is shown in Figure\,\ref{fig:fsr_plus_vdsr_network_skizze}\,(a).

For the network architecture, we use a custom Tensorflow~\cite{tensorflow2015-whitepaper} and Keras~\cite{chollet2015keras} implementation close to the original work in~\cite{Kim2016}.
For this implementation, as defined in the original paper, 20 convolutional layers with subsequent Rectified Linear Unit (ReLU) layers are used with $3{\times}3{\times}64$ filter kernels to calculate the residual image. This results in feature maps of the size $W{\times} H{\times}64$, with W and H representing the width and the height of the input image, respectively.
Zero-padding is applied at the boundaries for the convolutions. During training, the bicubically upscaled images are separated into patches with a size of $41{\times}41$ pixels and compared against the corresponding patches from the original data.
Therewith, the weights are adapted using the Adam optimizer~\cite{kingma2015} which showed superior performance than the stochastic gradient decent (SGD) method used for the original VDSR. As a loss function, the euclidean distance loss is taken.
Using Adam optimizer also requires to adapt the learning rate, which we initially set to 0.0001 and decrease by a factor of 10 after every 10-th epoch.
In total, the network is trained for 30 epochs. To increase the number of patches, and thus avoid overfitting during training, data augmentation is used by flipping and rotating the patches before feeding into the network. We further choose a batch size of 64 and a gradient clip value of 0.1.

\subsection{\mbox{VDSR} and proposed \mbox{VDSR-QS} for Quarter Sampling Sensors}

For the case of the quarter sampling sensor, a very similar processing chain can be followed. We describe the non-regular sub-sampling of the quarter sampling sensor as an element-wise multiplication of the reference image $f_{ij}$ with a binary mask $b_{ij}$. At this point, the sampled image $ (f_{ij} {\cdot} b_{ij})$ contains as many non-zero entries as pixels of the quarter sampling sensor. In order to reconstruct the missing pixels on the high-resolution grid, we use FSR \cite{Seiler2015} leading to a first reconstruction which we also call $\hat{f}_{ij}$ due to the strong similarity with the bicubically upscaled image in the previous section. This reconstructed image is now fed into the VDSR network which  we retrain for this data. This processing chain is  illustrated in Figure\,\ref{fig:fsr_plus_vdsr_network_skizze}\,(b). Here, an interesting observation can be made: Since FSR leaves the pixels that were actually measured with the quarter sampling sensor untouched, we know that those pixels where the condition $b_{ij}=1$ is true are identical to the pixel values in the reference image. It is therefore reasonable to set the residual at the output of the neural network to zero for those positions. The final image than reads 
\begin{align*}
	\tilde{f}^*_{ij} = \hat{f}_{ij} + r_{ij}\cdot(1-b_{ij}).
\end{align*}
This adaptation of VDSR is named \mbox{VDSR-QS} and is highlighted with red color in Figure\,\ref{fig:fsr_plus_vdsr_network_skizze}\,(b). Due to the differences, the \mbox{VDSR-QS} network is also trained independently from scratch.

Since the \mbox{VDSR} network and \mbox{VDSR-QS} network are mostly similar, the same set of hyper-parameters are used. The only exception is the learning rate for \mbox{VDSR-QS}. Since 25\% of the entries of the residual $r_{ij}\cdot(1-b_{ij})$ are set to zero, the loss function is  25\% smaller in average. To compensate for this, we multiply the learning rate in \mbox{VDSR-QS} by a factor of 4/3. In terms of the quarter sampling mask, we use the optimized quarter sampling mask from \cite{Grosche2018} because it shows an improved reconstruction quality for $\hat{f}_{ij}$. This mask is of size $32{\times}32$ pixels and is repeated periodically until the respective reference image is covered. Conveniently, such periodicity is beneficial for a future hardware implementation.

\subsection{Novel Data Augmentation for Quarter Sampling Sensor}

\label{sec:data_augmentation}
\begin{figure}[t]
	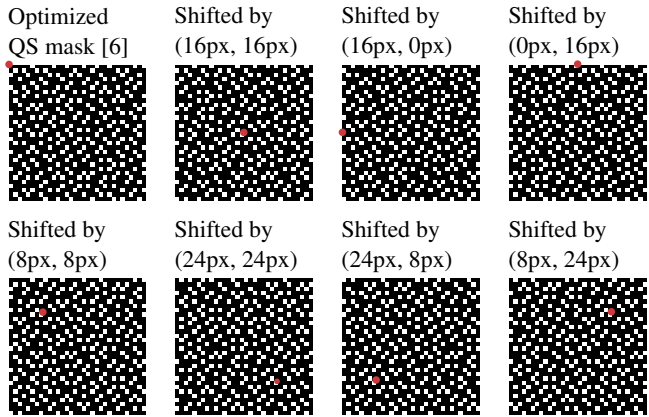
	\caption{Quarter sampling (QS) mask from \cite{Grosche2018} and its shifted variants used for sampling. Each mask is repeated periodically until the respective reference image is covered completely. The red dots indicate the top left corner of the non-shifted mask.}
	\label{fig:shifted_masks}
		\vspace*{-2mm}
\end{figure}

Intriguingly, using the quarter sampling sensor allows for a novel dimension for the data augmentation. From the available training images, more training data can be created by performing measurements with different sampling masks leading to different reconstructions $\hat{f}_{ij}$. To achieve this, the sampling mask is shifted by several pixels relative to the reference image. The different inputs $\hat{f}_{ij}$ to the neural network effectively increase the amount of training data. For the used sampling mask from \cite{Grosche2018}, a total of 1024 unique shifts are conceivable. We use the first, only the first two, only  the first four as well as all shifted masks shown in Figure\,\ref{fig:shifted_masks} were used in independent trainings. While 7 further doublings of the number of used masks are thinkable this would also lead to an increase of the computation time for the training phase in the same manner and is therefore omitted. 

\section{SIMULATIONS AND RESULTS}

\label{sec:simulation_and_results}
In this section, we evaluate the performance of the VDSR-like networks for image measurements using a quarter sampling sensor. First, the influence of the adapted network layout in \mbox{VDSR-QS} is evaluated relative to \mbox{VDSR} and  the influence from the novel data augmentation is shown. Second, the combination of both proposals is used to compare the reconstruction quality with the case of the low-resolution sensor.

In terms of the initial upscaling/reconstruction algorithms, we use bicubic upscaling and FSR with the same parameters as in \cite{Seiler2015}.
For the neural networks, we use our own custom TensorFlow \cite{tensorflow2015-whitepaper} implementation which was tested to show comparable results with respect to the original work in \cite{Kim2016}.
One training is performed for the low-resolution sensor. Please note, that this case is slightly different from the case investigated in \cite{Kim2016} as they concentrate on a different downscaling filter not accessible in a low-resolution sensor. In case of the quarter sampling sensor, we performed eight separate trainings. A factor of four in the number of training arises from the different amounts of data augmentation as discussed in Section\,\ref{sec:data_augmentation} and another factor of two arises from training the networks for the \mbox{VDSR-QS}, too.
In terms of the training data, we use the image Set 291 as in \cite{Kim2016} consisting of 291 images of various natural content. %
We simulate the monochrome low-resolution and quarter sampling sensors by converting all color images to grayscale with 8 bit depth.

To evaluate the quality of the resulting images, we calculate the mean PSNR and the mean structural similarity (SSIM) \cite{Wang2004} for the images of the Urban 100 dataset from \cite{Huang2015}. This dataset consists of various images of urban architecture. The content is of special interest as it is shows many of the structures where a low-resolution sensor may fail due to aliasing. 

\begin{figure}[t]
	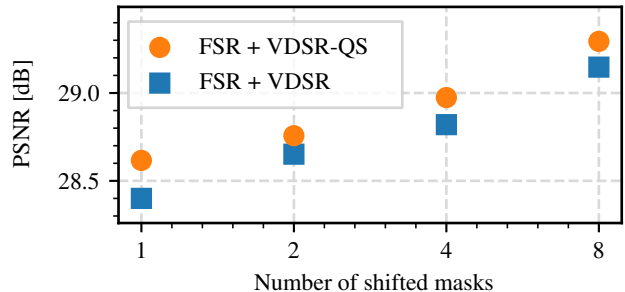
	\caption{Average PSNR for the Urban 100 dataset measured with the quarter sampling sensor. The cases of the \mbox{VDSR} (blue squares) and the \mbox{VDSR-QS} (orange circles) are shown. On the horizontal axis, the number of shifted masks used for data augmentation is given.}
	\label{fig:results_effect_aug_and_adapt}
		\vspace*{-2mm}
\end{figure}
Figure\,\ref{fig:results_effect_aug_and_adapt} shows the reconstruction quality in terms of PSNR using the quarter sampling sensor together with FSR and VDSR (blue squares) as well as \mbox{VDSR-QS} (orange circles). Four different amounts of data augmentation by shifting the quarter sampling mask were used as shown on the horizontal axis.
With an increasing amount of data augmentation, the PSNR increases by roughly \SI[retain-explicit-plus]{+0.2}{dB} per doubling, leading to  an increase of \SI[retain-explicit-plus]{+0.74}{dB} in total for the  VDSR. Interestingly, no saturation is yet visible for the shown number of used masks. 
Using the \mbox{VDSR-QS} leads to an additional gain of up to \SI[retain-explicit-plus]{+0.21}{dB}. %
Here, a similar increase  is observed across all amounts of data augmentation.
Overall, an increase of \SI[retain-explicit-plus]{+0.89}{dB} %
was achieved by incorporating the knowledge that the quarter sampling sensor was used and by exploiting the novel degree of freedom in the data augmentation.

With these findings, we next compare the reconstruction quality using the quarter sampling sensor and the low-resolution sensor, both used with the VDSR networks. In Table\,\ref{tab:results_BIC_vs_FSR_each_with_VDSR} the image quality in terms of PSNR and SSIM is shown. For completeness, the intermediate image quality after the upscaling/reconstruction using bicubic interpolation/FSR is also provided. For the quarter sampling sensor, the data is given for the case of the \mbox{VDSR} (no adaptation, no novel data augmentation) and the  \mbox{VDSR-QS} with 8-fold data augmentation using all masks from Figure\,\ref{fig:shifted_masks} which leads to the highest PSNR and highest SSIM.

\begin{table}[t]
	\caption{Image quality in terms of average PSNR in dB and SSIM using the low-resolution (LR) sensor and the quarter sampling (QS) sensor for the Urban 100 image dataset. Besides providing PSNR/SSIM without using the VDSR networks, the PSNR/SSIM at the output of VDSR and VDSR-QS is given. For the latter, the result with the novel 8-fold data augmentation (DA) is given. Bold font indicates best PSNR/SSIM.}
	\label{tab:results_BIC_vs_FSR_each_with_VDSR}
	\centering
	\begin{tabular}{l|c|c}
		                                               &   PSNR [dB]    & SSIM\\ \hline
		Low-resolution sensor                          &                & \\
		\,\,\,\,\, BIC  upscaling                      &     25.67      & 0.8818 \\
		\,\,\,\,\, BIC + VDSR \cite{Kim2016}           &     28.62      & 0.9265 \\ \hline
		Quarter sampling sensor                        &                & \\
		\,\,\,\,\, FSR \cite{Seiler2015}               &     27.08      & 0.9116 \\
		\,\,\,\,\, FSR + \mbox{VDSR}                   &     28.40      & 0.9281 \\
		\,\,\,\,\, FSR + \mbox{VDSR-QS}    + 8-fold DA & \textbf{29.29} & \textbf{0.9382} 
	\end{tabular}
\vspace*{-3mm}
\end{table}

From Table\,\ref{tab:results_BIC_vs_FSR_each_with_VDSR}, we find that the reconstruction quality using the quarter sampling sensor with the \mbox{VDSR} without adaptations is  \SI{0.22}{dB} lower than using the low-resolution sensor with VDSR. However, using \mbox{VDSR-QS} and the additional 8-fold data augmentation, the quarter sampling sensor achieves a image quality of \SI{29.29}{dB} in terms of PSNR, being \SI[retain-explicit-plus]{+0.67}{dB} %
higher than for the VDSR with the low-resolution sensor. Compared to the reconstruction using only the FSR, the \mbox{VDSR-QS} together with an 8-fold data augmentation gains more than \SI[retain-explicit-plus]{+2.2}{dB}.
The SSIM values verify all these findings. Overall, a SSIM gain of \num[retain-explicit-plus]{+0.0266} compared to using FSR is achieved using \mbox{VDSR-QS} and the 8-fold data augmentation.

\begin{figure}[t!]
	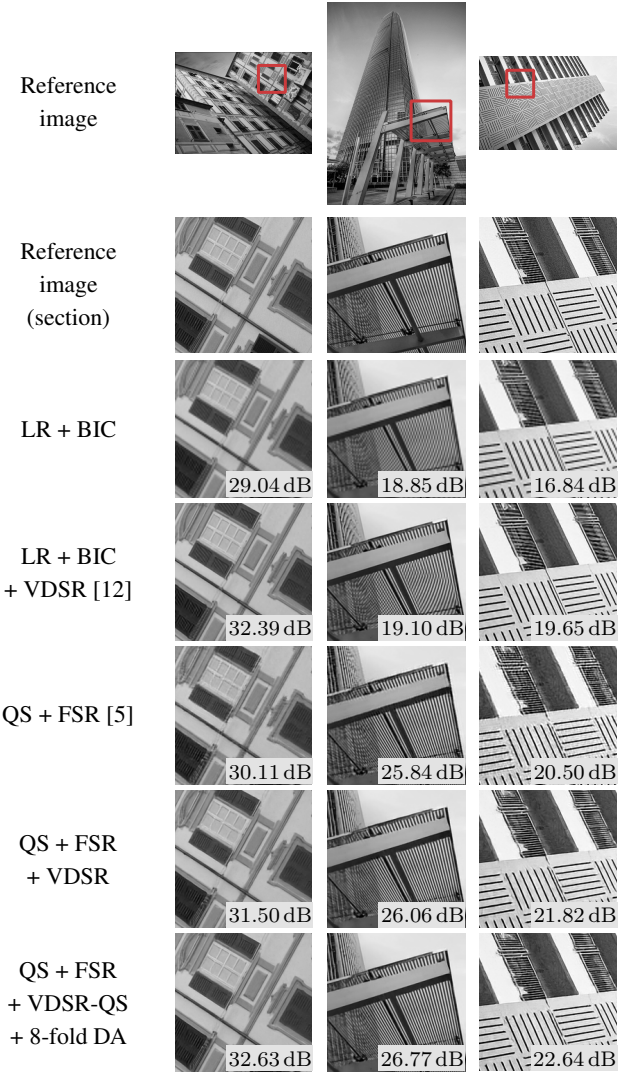
	\caption{Visual examples from sections of images of the Urban 100 dataset. While the low-resolution (LR) sensor leads to severe aliasing, the \mbox{VDSR-QS} combined with our proposed 8-fold data augmentation can recover fine high-frequency details from the measurement of the quarter sampling (QS) sensor. The insets provide PSNR values of the shown sections. \textit{(Please pay attention, additional aliasing may be caused by printing or scaling. Best to be viewed enlarged on a monitor.)}}
		\vspace*{-3mm}
	\label{fig:results_visual}
	\vspace*{-2mm}
\end{figure}
In Figure\,\ref{fig:results_visual}, we show visual examples for three sections of images from the Urban 100 dataset. Regions with high frequency content are affected from severe aliasing in case of the low-resolution sensor regardless of using VDSR or not. Other than that, using the quarter sampling sensor together with FSR and \mbox{VDSR} or \mbox{VDSR-QS} increases the image quality. This is especially visible for high-frequency content where the aliasing may even change the angle of parallel lines for the low-resolution sensor as can be seen in the second image. Compared to the reconstruction with the FSR, the ringing-like artifacts, e.g., occurring around some of the edges, are significantly reduced, while the correct high-frequency content is preserved.

\vspace*{-1mm}
\section{CONCLUSION}
\label{sec:conclusion}
In this paper, we transfer the concepts of Very Deep Super Resolution network (VDSR) to the special case of quarter sampling enhance the reconstruction quality of image data acquired with a quarter sampling sensor.
In doing so, we propose an adapted version of VDSR called \mbox{VDSR-QS} to incorporate the special property of the quarter sampling measurements that the exact value of some of the high-resolution pixels are known. Moreover, we examine a novel degree of freedom in the data augmentation only possible for the case of quarter sampling. Combining the \mbox{VDSR-QS} with such 8-fold data augmentation,  we successfully increase the reconstruction quality of images measured with a quarter sampling sensor by \SI[retain-explicit-plus]{+2.2}{dB}. Compared to a low-resolution sensor and  VDSR, we gain \SI[retain-explicit-plus]{+0.67}{dB}.
Moreover, we provide visual examples showing that aliasing occurring from a measurement with the low-resolution sensor is not present for the quarter sampling sensor and the artifacts usually occurring in quarter sampling measurement reconstructed with FSR are significantly reduced.

For future work, we would like to transfer the advantages of \mbox{VDSR-QS} to other non-regular measurement scenarios such as three-quarter sampling, where randomly oriented L-shaped pixels are used resulting in higher reconstruction quality \cite{Seiler2018}. Even though, similar to the low-resolution sensor, a low-pass filter is effectively applied prior to the measurement, the advantages of the quarter sampling are kept for the three-quarter sampling. However, due to the filtering, adaptations such as in \mbox{VDSR-QS} are not straightforward.

\vspace*{-2mm}
\section{ACKNOWLEDGMENT}
\vspace*{-2mm}
We gratefully acknowledge that this work has been supported by the Deutsche Forschungsgemeinschaft (DFG) under contract number KA 926/5-3.

\clearpage

\bibliographystyle{IEEEbib}
\bibliography{literatur_jabref}

\begin{thebibliography}{10}

\bibitem{Schoberl2011}
Michael Sch{\"o}berl, J{\"u}rgen Seiler, Siegfried Foessel, and Andr{\'e} Kaup,
\newblock ``Increasing imaging resolution by covering your sensor,''
\newblock in {\em Proc. 18th {IEEE} International Conference on Image
  Processing}, Brussels, Sept. 2011, pp. 1897--1900.

\bibitem{Dippe1985}
Mark A.~Z. Dipp{\'{e}} and Erling~Henry Wold,
\newblock ``Antialiasing through stochastic sampling,''
\newblock in {\em Proc. 12th Annual Conference on Computer Graphics and
  Interactive Techniques}, New York, July 1985, pp. 69--78.

\bibitem{Hennenfent2007}
Gilles Hennenfent and Felix~J. Herrmann,
\newblock ``Irregular sampling: from aliasing to noise,''
\newblock in {\em Proc. 69th EAGE Conference and Exhibition}, London, June
  2007, pp. cp--27--00063.

\bibitem{Maeda2009}
Yui Maeda and Junichi Akita,
\newblock ``A {CMOS} image sensor with pseudorandom pixel placement for clear
  imaging,''
\newblock in {\em Proc. International Symposium on Intelligent Signal
  Processing and Communication Systems}, Kanazawa, Dec. 2009, pp. 367--370.

\bibitem{Seiler2015}
J{\"u}rgen Seiler, Markus Jonscher, Michael Sch{\"o}berl, and Andr{\'e} Kaup,
\newblock ``Resampling images to a regular grid from a non-regular subset of
  pixel positions using frequency selective reconstruction,''
\newblock {\em {IEEE} Transactions on Image Processing}, vol. 24, no. 11, pp.
  4540--4555, Nov. 2015.

\bibitem{Grosche2018}
Simon Grosche, J{\"u}rgen Seiler, and Andr{\'e} Kaup,
\newblock ``Iterative optimization of quarter sampling masks for non-regular
  sampling sensors,''
\newblock in {\em Proc. International Conference on Image Processing 2018},
  Athens, Oct. 2018, pp. 26--30.

\bibitem{Herraiz2008}
Joaquin~Lopez Herraiz, Samuel Espana, Esther Vicente, Elena Herranz, Manuel
  Desco, Juan~Jose Vaquero, and Jose Udias,
\newblock ``Frequency selective signal extrapolation for compensation of
  missing data in sinograms,''
\newblock in {\em Proc. {IEEE} Nuclear Science Symposium Conference Record},
  Dresden, Oct. 2008, pp. 4299--4302.

\bibitem{Stehle2006}
Thomas Stehle,
\newblock ``Removal of specular reflections in endoscopic images,''
\newblock {\em Acta Polytechnica}, vol. 46, no. 4, pp. 32, 2006.

\bibitem{Seiler2009}
J{\"u}rgen Seiler and Andr{\'e} Kaup,
\newblock ``Multiple selection extrapolation for improved spatial error
  concealment,''
\newblock in {\em Proc. {IEEE} International Workshop on Multimedia Signal
  Processing}, Rio de Janeiro, Oct. 2009, pp. 1--6.

\bibitem{Yang2010}
Jianchao Yang, John Wright, Thomas~S Huang, and Yi~Ma,
\newblock ``Image super-resolution via sparse representation,''
\newblock {\em {IEEE} Transactions on Image Processing}, vol. 19, no. 11, pp.
  2861--2873, nov 2010.

\bibitem{Dong2016}
Chao Dong, Chen~Change Loy, Kaiming He, and Xiaoou Tang,
\newblock ``Image super-resolution using deep convolutional networks,''
\newblock {\em {IEEE} Transactions on Pattern Analysis and Machine
  Intelligence}, vol. 38, no. 2, pp. 295--307, feb 2016.

\bibitem{Kim2016}
Jiwon Kim, Jung~Kwon Lee, and Kyoung~Mu Lee,
\newblock ``Accurate image super-resolution using very deep convolutional
  networks,''
\newblock in {\em Proc. {IEEE} Conference on Computer Vision and Pattern
  Recognition}, Las Vegas, June 2016, pp. 1646--1654.

\bibitem{tensorflow2015-whitepaper}
Mart\'{\i}n Abadi, Ashish Agarwal, Paul Barham, et~al.,
\newblock ``{TensorFlow}: Large-scale machine learning on heterogeneous
  systems,'' 2015,
\newblock Software available from \url{tensorflow.org}.

\bibitem{chollet2015keras}
Fran\c{c}ois Chollet et~al.,
\newblock ``Keras,'' \url{https://keras.io}, 2015.

\bibitem{kingma2015}
Diederik~P. Kingma and Jimmy Ba,
\newblock ``Adam: A method for stochastic optimization,''
\newblock in {\em Proc. International Conference on Learning Representations},
  San Diego, May 2015.

\bibitem{Wang2004}
Zhou Wang, Alan~Conrad Bovik, Hamid~Rahim Sheikh, and Eero~P. Simoncelli,
\newblock ``Image quality assessment: From error visibility to structural
  similarity,''
\newblock {\em {IEEE} Transactions on Image Processing}, vol. 13, no. 4, pp.
  600--612, Apr. 2004.

\bibitem{Huang2015}
Jia-Bin Huang, Abhishek Singh, and Narendra Ahuja,
\newblock ``Single image super-resolution from transformed self-exemplars,''
\newblock in {\em Proc. {IEEE} Conference on Computer Vision and Pattern
  Recognition}, Boston, June 2015, pp. 5197--5206.

\bibitem{Seiler2018}
J{\"u}rgen Seiler, Markus Jonscher, Thomas Ussmueller, and Andr{\'e} Kaup,
\newblock ``Increasing imaging resolution by non-regular sampling and joint
  sparse deconvolution and extrapolation,''
\newblock {\em {IEEE} Transactions on Circuits and Systems for Video
  Technology}, vol. 29, no. 2, pp. 308--322, Feb. 2019.

\end{thebibliography}

\end{document}